\documentstyle[11pt]{article}
\def\beq{\begin{eqnarray}}
\def\eeq{\end{eqnarray}}

\def\al{\alpha}
\def\be{\beta}

\def\ga{\gamma}

\def\ze{\zeta}

\def\la{\lambda}
\def\na{\nabla}

\begin{document}

\hfill Preprint number

\hfill Date


\vskip 3mm

\begin{center}
{\Large {\bf Renormalization group study of the higher derivative }}
\vskip 
2mm {\Large {\bf conformal scalar model}}

\vskip 8mm

{\bf J. A. de Barros $^{a}$} \footnote{%
Electronic address: acacio@fisica.ufjf.br},$\,\,\,\,\,\,$ {\bf I. L. Shapiro 
$^{a,b}$} \footnote{%
Electronic address: shapiro@fisica.ufjf.br} \vskip 4mm

{\sl a) Departamento de Fisica -- ICE, Universidade Federal de Juiz de Fora,
Juiz de Fora, 36036-330, MG, Brazil}

{\sl b) Tomsk State Pedagogical Institute, Tomsk, 634041, Russia}

\vskip 8mm
\end{center}

\noindent {\large {\sl Abstract.}}$\,\,\,${\it The second alternative
conformal limit of the recently proposed general higher derivative dilaton
quantum theory in curved spacetime is explored. In this version of the
theory the dilaton is transformed, along with the metric, to provide the
conformal invariance of the classical action. We find the corresponding
quantum theory to be renormalizable at one loop, and the renormalization
constants for the dimensionless parameters are explicitly shown to be
universal for an arbitrary parametrization of the quantum field. The
renormalization group equations indicate an asymptotic freedom in the IR
limit. In this respect the theory is similar to the well-known model based
on the anomaly-induced effective action.}

\vskip 12mm 
\noindent {\large {\bf Introduction}} \vskip 2mm

The gravitational effective action, generated by the trace anomaly of the
conformal invariant matter in curved spacetime, is of considerable interest
due to numerous physical applications (see \cite{20let} for the review) and
also because of the hope to use such an action as an insight for the theory
of quantum gravity \cite{odsh,anmo,desc,cosh,deser}. The first solution for
the effective action has been obtained by Reigert \cite{rei} and by Fradkin
and Tseytlin \cite{frts} (see also \cite{buodsh,bugufo,dow}). The equation
solved in \cite{rei,frts} links the effective action with the trace anomaly
of the Energy-Momentum tensor 
\begin{equation}
-{\frac{2}{\sqrt{-g}}}\,g_{\mu \nu }{\frac{\delta \Gamma }{{\delta g_{\mu
\nu }}}}=T_{\mu }^{\mu }.  \label{effact}
\end{equation}
\noindent The above equation remains unaltered if one changes the solution
according to $\Gamma \rightarrow \Gamma +S_{c}$, where
$S_{c}(g_{\mu \nu })$
is an arbitrary conformal invariant functional. Therefore this equation
doesn't define $\Gamma (g_{\mu \nu })$ completely, and in fact the most
complicated part of the effective action remains hidden in 
$S_{c}(g_{\mu \nu})$. 
As a result  the nonconformal part of the effective action is ambiguous, and
the particular version derived in \cite{rei,frts} fails to pass the test
based on the calculation of the three point functions for the gravitational
field \cite{desc,osbpet,erdosb}. Consequently, it is important to find a
solution for (\ref{effact}) which assumes such a test, but this problem has
not
been solved yet \cite{deser}. Another discrepancy arises if one compares
the solution of \cite{rei,frts} with the result of direct calculations of
the effective action performed in \cite{vilk}. The effective action for the
conformal scalar field \cite{vilk} is a nonlocal functional -- as well as
the solution of \cite{rei,frts} -- but the nonlocalities are related with
the Green functions of the second order conformal operator 
\begin{equation}
\Delta _{2}={\Box }-\frac{1}{6}\,R,  \label{2oper}
\end{equation}
while in \cite{rei,frts} they are related with the Green function of the
fourth order conformal operator 
\begin{equation}
\Delta _{4}={\Box }^{2}+2\,R^{\mu \nu }\,\nabla _{\mu }\,\nabla _{\nu }-%
\frac{2}{3}\,R\,\Box -\frac{1}{3}\,\nabla ^{\mu }\,R\,\nabla _{\mu }.
\label{4oper}
\end{equation}
Indeed, any two solutions of (\ref{effact}) differ by the conformally
invariant functional, which is supposed to be nonlocal and probably very
complicated. At the same time, in many known cases the nonlocality can be
removed by
introducing the auxiliary scalar fields, while all the symmetries
of the theory are preserved. Therefore, it is worth to study a possible
forms of conformal invariant metric-dilaton models and relations between
them.

In previous papers \cite{a,eli} we have studied the renormalization and the
renormalization group of the general high derivative dilaton model where we
were especially interested in its conformal limit associated with the $%
\Delta _{4}$ operator (\ref{4oper}). Here we will be mainly concerned and
try to complete the similar program about the second (alternative) conformal
invariant limit of the same general theory \cite{eli}. This alternative
conformal model is based on the second order operator (\ref{2oper}) and in
this respect it is closer to the desirable form of the anomaly-induced
effective action for gravity \cite{desc,vilk,deser}. Our conformal theory is
a direct (and quite simple) generalization of the one proposed some
years ago
by Antoniadis, Iliopoulos and Tomaras \cite{anilto}, and differs from it by
the reparametrization of the scalar field \cite{conf}.

The paper is organized in the following way. In the next section we write
down the action of the general high derivative dilaton model and of the
both
of its conformal limits. We ascertain that in $D=4$ the difference between
these two conformal invariant theories is the low of conformal
transformation for the dilaton. Moreover, we try to generalize both
theories
to $n$ dimensions, $n$ different from 4, and show that in any $n\neq 4$ the
transformation low for the dilaton in two theories can be equal. Some
special transformation properties of the second conformal model are also
considered, and the higher derivative generalization of the conformal
duality \cite{bek,conf,duco} is formulated. In section 3 the one-loop
divergences for the second conformal model are derived and their
parametrization dependence is studied. In the fourth section
we consider the
renormalization group equations for the model. The last section contains 
conclusion and some suggestions for the future study.

\vskip 6mm 
\noindent {\large {\bf 2. General high derivative scalar model and two of
its conformal limits.}} \vskip 2mm

The action of the general high derivative scalar model has the form \cite
{eli} 
\[
S=\int d^{4}x\sqrt{-g}\{b_{1}\,(\Box \varphi )^{2}+b_{2}\left( \nabla
\varphi \right) ^{2}\Box \varphi +b_{3}(\nabla \varphi )^{4}+b_{4}(\nabla
\varphi )^{2}+b_{5}+c_{1}R(\nabla \varphi )^{2}+
\]
\begin{equation}
+c_{2}R^{\mu \nu }\,\partial _{\mu }\varphi \,\partial _{\nu }\varphi
+c_{3}R\Box \varphi +a_{1}R_{\mu \nu \alpha \beta }^{2}+a_{2}R_{\mu \nu
}^{2}+a_{3}R^{2}+a_{4}R\}+(\mbox{surface terms}).  \label{gener}
\end{equation}
Here the generalized couplings $a,b,c$ are some functions of the scalar
field $\varphi $. $a,b,$ and $c$  are dimensionless, except for $b_{4}$,
$b_{5}$ and $a_{4}$, for which we have: $b_{4}(\varphi )\sim m^{2}$,
$b_{5}(\varphi )\sim m^{4}$, $a_{4}(\varphi )\sim m^{2}$. All other
possible
terms, with an appropriate dimension which can be included into the action,
differ from the ones which are already present by surface terms only \cite
{eli}. We also used the notation 
$g^{\mu \nu }\,\partial _{\mu}\varphi 
\,\partial _{\nu }\varphi  =(\nabla \varphi )^{2}$.

The first conformal limit of the metric-dilaton model is related with the
solution of \cite{rei,frts}. Such a conformal model has been formulated in 
\cite{a,eli}. The action of the model is 
\begin{equation}
S_{c}^{(1)}=\int d^{4}x\sqrt{-g}\,\left\{ \;{\frac{1}{{2}}}\,f(\varphi
)\,\varphi \Delta _{4}\varphi +q(\varphi )\,C_{\mu \nu \alpha \beta
}\,C^{\mu \nu \alpha \beta }+p(\varphi )\,(\nabla \varphi )^{4}+v\,E\right\}
,  \label{conf1}
\end{equation}
where $f,q,p$ are some differentiable functions of $\varphi $,
$\,v=const.$, 
and $C_{\alpha \beta \mu \nu }$ is the Weyl tensor. The square of this
tensor will be denoted in what follows as $C_{\mu \nu \alpha \beta }\,C^{\mu
\nu \alpha \beta }=C^{2}$. The action includes the Gauss-Bonnet topological
term $\int E,\,$ where
$E=R_{\alpha \beta \mu \nu }^{2}-4R_{\alpha \beta}^{2}+R^{2}$.
This term is a full derivative in four dimensions, but in $%
n\neq 4$ it cannot be ajusted to be conformally invariant, while
$C^{2}$ can.

As was discussed in \cite{a}, the action (\ref{conf1}) includes the term
$\,\varphi \Delta _{4}\varphi \,$ which is a necessary component of the
solution for the anomaly-induced effective action \cite{rei}, as far as we
want to write it in a local form. In \cite{eli} it was shown that, at one
loop, the generic theory (\ref{conf1}) is multiplicatively renormalizable.
Moreover, the one-loop divergences has been obtained. In accordance with the
general theorem \cite{buch,book} the one-loop divergences are conformal
invariant if the initial model had such invariance. In the calculations of 
\cite{eli} a background field method and Schwinger-DeWitt formalism in its
modern form were used, and only the scalar field was considered as a
quantum
field, while the metric was regarded as a purely classical background. In
this way, it was found that for some special choice of the conformal model,
that is for some particular form of the functions $f,q,p$, the theory is
one-loop finite (in the dimensional regularization) and, therefore, it is
free from the conformal anomalies.

The action (\ref{conf1}) may be written as a particular case of
(\ref{gener}) --
with special conformal constraints which have the form (below we shall
frequently abandon the argument $\phi $) 
\begin{equation}
a_{1}=-\frac{1}{2}\,a_{2}={3}a_{3}=q(\varphi
),\,\,\,\,\,\,\,\,b_{1}=f^{\prime }(\varphi )\varphi +
f(\varphi )=\frac{3}{2}%
\,c_{1}=-\frac{1}{2}\,c_{2},\,\,\,\,\,\,\,\,b_{2}=b_{1}^{\prime
},\,\,\,\,\,\,\,\,b_{3}=p(\varphi )  \label{constr}
\end{equation}
while the rest of the generalized couplings $a_{4}(\varphi ),b_{4}(\varphi
),b_{5}(\varphi ),c_{3}(\varphi )$ are equal to zero, and we use ``$^{\prime
}$'' to denote differentiation with respect to $\varphi .$ The one-loop
finite power-form solutions for $b_{1,2,3}$ found in \cite{eli} have the
form (with accuracy to an overall constant): 
\begin{equation}
b_{1}=1,\;\;\;\;\;b_{2}=0,\;\;\;\;\;b_{3}=0,  \label{finsol1}
\end{equation}
\begin{equation}
b_{1}=1,\ \ \ \ \ \ b_{2}={b_{1}}^{\prime }=0,\ \ \ \ \ \ b_{3}={3}/{5}
\;(\varphi -\varphi _{0})^{-2},\;\;\;\;\;\varphi _{0}=\mbox{const.},
\label{finsol2}
\end{equation}
\begin{equation}
b_{1}=(\varphi -\varphi _{0})^{-2},\ \ \ \ b_{2}={b_{1}}^{\prime
}=-2(\varphi -\varphi _{0})^{-3},\ \ \ \ b_{3}=(\varphi -\varphi _{0})^{-4}.
\label{finsol1-1}
\end{equation}
In the second solution (\ref{finsol2}) $b_{3}$ can be generalized to
$b_{3}=F^{-1}(\varphi -\varphi _{0})$ where the function $F(p)$ is the
solution of the differential equation $p^{\prime \prime }-10p^{2}=0$. This
solution is independent on the first one (\ref{finsol1}), but one can easily
see that the third one (\ref{finsol1-1}) coincides with (\ref{finsol1})
after a nonpolinomial reparametrization $\varphi \rightarrow \ln \varphi$
in the former. Thus, we have only two independent solutions. This example
shows the importance of tracing an arbitrary parametrization of the
background field in the sigma-model type theories in $n=4$. The
parametrization of the quantum field is also important, since it can modify
the renormalization constants and therefore the beta-functions and scaling
behaviour.
A comprehensive formal study of the parametrization dependence in
quantum field theory can be found in \cite{param}. However, explicit loop
calculations in an arbitrary parametrization of the quantum field should
also be interesting. In the next section we investigate in details the
parametrization dependence for the one-loop divergences of the second
conformal high derivative scalar model (\ref{conf2}).

The model (\ref{conf1}) is invariant under a local conformal
transformation of the metric only
${\bar{g}}_{\mu \nu }=g_{\mu \nu}\,e^{2\sigma (x)},$ while the dilaton
is not transformed. However, in (\ref
{gener}) the transformation rule for the dilaton is not fixed, and that is
why there exists the second conformal limit of (\ref{gener}) with the
nontrivial transformation rule for the dilaton \cite{conf}. 
\begin{equation}
S_{c}^{(2)}=\int d^{4}x\sqrt{-g}\left\{ vE+xC^{2}+y\left[ \frac{R}{3}-
\frac{{\Box }\zeta }{\zeta }+\frac{({\nabla }\zeta )^{2}}{2\zeta ^{2}}
\right] ^{2}+
\frac{3z\zeta }{\gamma }\left[ \frac{R}{3}-
\frac{{\Box }\zeta}{\zeta }+
\frac{(\nabla \zeta )^{2}}{2\zeta ^{2}}\right] -
\frac{u\zeta^2}{\gamma^2}\right\},
\label{conf2}
\end{equation}
where $\zeta =\zeta (\phi )$ is the function of the scalar field
$\phi $,
$\gamma $ is (dimensional) Newton's constant, and $x,y,z,u$ are some
dimensionless constants. The last action is invariant under the
transformation which includes an arbitrary reparametrization of the scalar
field and the special conformal transformation of the metric 
\begin{equation}
{\bar{\phi}}={\bar{\phi}}(\phi ),\,\,\,\,\,\,\,\,\,\,\,\,\,\,
{\bar{g}}_{\mu\nu }=
g_{\mu \nu }\,\left[ \frac{\zeta ({\phi })}{\zeta ({\bar{\phi}}(\phi ))}
\right] .  \label{trans}
\end{equation}
One has to notice that the reparametrization of the scalar field alone,
without the transformation of the metric, doesn't change neither the
form of
the action (\ref{conf2}) nor the values of the constants
$\gamma ,v,x,y,z,u$.
It only changes the form of the function $\zeta (\phi )$. In this way one
can cast the action (\ref{conf2}) into the form originally proposed by
Antoniadis, Iliopoulos and Tomaras \cite{anilto} 
\begin{equation}
S_{c}=\int d^{4}x\sqrt{-g}\;\left\{ v\,E+x\,C^{2}+y\,\left( \phi
^{-1}\,\Delta _{2}\,\phi \right) ^{2}-\frac{6z}{\ga}\,
\phi \,\Delta _{2}\,\phi
-u\,\phi ^{4}\right\}.   \label{conf3}
\end{equation}
In general, the new action -- with another parametrization of the scalar
field -- will possess the same symmetry (\ref{trans}) as the original one.
Therefore the reparametrization of the dilaton in (\ref{conf2}) transfers
one such theory into another, with different form of $\zeta (\phi )$. The
same result can be achieved by doing the conformal transform of the metric
in (\ref{trans}) without affecting the dilaton. Moreover, the conformal
transformation of the metric can do what reparametrization of the scalar
cannot -- it can make the new $\zeta $ constant. As a result one arrives
at the familiar action for the higher derivative gravity (see (\cite
{stelle,book} and references therein) 
\begin{equation}
S_{HD}=
\int d^{4}x\sqrt{-{\bar{g}}}\;\left\{ vE+xC^{2}+y{\bar{R}}^{2}+
\frac{z}{3\gamma }{\bar{R}}-\frac{u}{\gamma ^{4}}\right\} ,  \label{high}
\end{equation}
with all $v,x,y,z,u,\gamma $ constants. Thus, the transformation properties
of (\ref{conf2}) are quite similar to the ones of the second derivative
conformal metric-dilaton theory considered in \cite{conf}. The difference is
that the second derivative model is conformally equivalent to the Einstein
gravity, whereas the fourth derivative model (\ref{conf2}) to the higher
derivative gravity (\ref{high}). In both cases an extra conformal symmetry
''eats'' the extra scalar field. Of course the second derivative sector of
(\ref{conf2}) reproduces the model of \cite{conf} literally.

Let us mention that one can easily formulate the conformal duality, which
takes place in the second derivative metric-dilaton models \cite
{bek,conf,duco}, for its fourth derivative cousin\footnote{%
Another form of conformal duality takes place in some purely metric
theories
with action $\int d^{4}x\sqrt{-g}L(R)$ \cite{hajo}. Such models can be
reduced to the second derivative metric-dilaton theory \cite{barrow}. The
underlying metric-dilaton model is associated with the operator
(\ref{2oper})
and in this sense can be close to the one considered in \cite{bek,conf}.}.
To see this we denote the action (\ref{conf2}) as $S_{\zeta (\phi
);v,x,y,z,u}$ and, following the method of \cite{conf}, construct a sum 
\begin{equation}
S_{\zeta (\phi );v,x,y,z,u}\left( \phi ,g_{\mu \nu }\right) +
S_{\xi (\phi);v^{\prime },x^{\prime },y^{\prime },z^{\prime },
u^{\prime }}\left( \phi,g_{\mu \nu }\right) ,  \label{dual1}
\end{equation}
where $v^{\prime },x^{\prime },y^{\prime },z^{\prime },u^{\prime }$
are some
other constants and $\xi (\phi )$ -- some other function. If we perform the
transformation of the metric $g_{\mu \nu }={\bar{g}}_{\mu \nu }\,\Omega
(\phi )$ and take $\Omega =N(\phi )/\zeta (\phi )$, with a differentiable
function $N(\phi )$, then (\ref{dual1}) becomes 
\begin{equation}
S_{N(\phi );x,y,z,u}\left( \phi ,{\bar{g}}_{\mu \nu }\right) +
S_{\frac{N(\phi )\xi (\phi )}{\zeta (\phi )};x^{\prime },y^{\prime },
z^{\prime},
u^{\prime }}\left( \phi ,{\bar{g}}_{\mu \nu }\right) .  \label{dual2}
\end{equation}
In particular, for $\zeta =\phi ^{2}$ and $\xi =\kappa ^{2}=const$ we meet
the symmetry which is a direct generalization of the conformal duality of
$\,\,$ \cite{bek,conf,duco}: 
\begin{equation}
\phi \longleftrightarrow \frac{1}{\phi },\,\,\,\,\,\,\,\,\,\,\kappa
\longleftrightarrow \frac{1}{\kappa },\,\,\,\,\,\,\,\,\,\,(x,y,z,u)%
\longleftrightarrow (x^{\prime },y^{\prime },z^{\prime },u^{\prime})
,\,\,\,\,\,\,\,\,\,\,g_{\mu \nu }\longleftrightarrow {\bar{g}}_{\mu \nu }=
\frac{\zeta }{N}\,g_{\mu \nu }.  \label{dual3}
\end{equation}
One can formulate this dual symmetry in any spacetime dimension different
than $2$, as it was done in \cite{duco} with the second derivative 
metric-dilaton model.

As we have seen, there are two conformal limits in the general model (\ref
{gener}), the only difference between them is a different rule of
transformation for the dilaton. As far as in quantum domain we suppose to
use the dimensional regularization, it is significant that one can, in some
extent, formulate the conformal invariant models of both types in a
spacetime dimension different than $4$. To construct a version of
(\ref{conf1})
in arbitrary $n$ we have to use the $n$-dimensional generalization
of the fourth-order conformally-covariant self-adjoint operator
$\Delta _{4}$
(\ref{4oper}). The $\Delta _{4}$ operator was originally invented in an
arbitrary dimension \cite{pan} (see also \cite{branson,erd} for the wider
classes of conformal operators and \cite{branson2} for the recent
investigation of $\Delta _{4}$), where it has the form 
$$
\Delta _{4}^{(n)}={\Box }^{2}+\frac{4}{n-2}\,\left[R^{\mu \nu }\,
\nabla _{\mu }\,\nabla _{\nu}+\frac{1}{2}\,(\na^\mu R)\na_\mu \right] +
\frac{4n-n^2-8}{2(n-1)(n-2)}
\,\left[R\,\Box + (\na^\mu R)\na_\mu \right] +
$$
\begin{equation}
+(4-n)\,\left[\frac{1}{4(n-1)}(\Box R) + \frac{8-5n}{16(n-1)^2(n-2)}R^2 
+ \frac{1}{(n-2)^2}\,R_{\mu\nu}\,R^{\mu\nu} \right] .
\label{4opern}
\end{equation}
The last term in (\ref{conf1}) cannot be conformally continued to $n\neq 4$
and hence the corresponding action has the form 
\begin{equation}
S_{c}^{(1,n)}=\int d^{n}x\sqrt{-g}\,\left\{ \;{\frac{1}{{2}}}\,f
\varphi \Delta _{4}^{(n)}\varphi +q\,\varphi ^{2}\,C^{2}\right\} ,
\label{conf1n}
\end{equation}
with $f,q=const's$.
The transformation low is
$g_{\mu \nu }\rightarrow g_{\mu\nu }^{\prime }=
g_{\mu \nu }\,\exp (2\sigma ),\,\,\,\varphi \rightarrow
\varphi ^{\prime }=\varphi \,\exp (\frac{4-n}{2}\sigma )$. For $n=4$ the
above action is reduced to a particular case of (\ref{conf1}). Indeed there
is the possibility of an arbitrary reparametrization of the scalar (just as
in (\ref{conf1}), if we do so, then the form of the conformal
transformation should be
modified in the spirit of (\ref{trans})). We remark, that the pure Weyl
gravity is nonconformal in $n\neq 4$, and the scalar field $\varphi $ (with
a nontrivial transformation low) serves as a compensator which provides the
invariance of the $C^{2}$-term.

The $n$-dimensional form of (\ref{conf2}) can be easily written for the
Antoniadis, Iliopoulos and Tomaras theory (\ref{conf3}). As we already
mentioned above, this version differs from the general one (\ref{conf2}) by
the reparametrization of the scalar, so one can easily rewrite the result
for the case of (\ref{conf2}). In the $n$-dimensional case the conformal
invariant action is 
\begin{equation}
S_{c}=\int d^{n}x\sqrt{-g}\;\left\{ x\,\phi ^{\frac{2(n-4)}{n-2}%
}C^{2}+y\,\phi ^{\frac{2(n-4)}{n-2}}\,\left( \phi ^{-1}\,\Delta
_{2}^{(n)}\,\phi \right) ^{2}-\frac{6z}{\ga}\,\phi \,\Delta _{2}^{(n)}\,\phi
-u\,\phi ^{\frac{2n}{n-2}}\right\} ,  \label{conf2n}
\end{equation}
where the operator $\Delta _{2}^{(n)}$ and the transformation rules have
usual form (see, for example, \cite{book}) 
\begin{equation}
\Delta _{2}^{(n)}=\Box +\frac{n-2}{4(n-1)}\,R,\,\,\,\,\,\,\,\,\,g_{\mu \nu
}\rightarrow g_{\mu \nu }^{\prime }=g_{\mu \nu }\,e^{2\sigma
},\,\,\,\,\,\,\,\,\phi \rightarrow \phi ^{\prime }=\phi \,e^{\frac{n-2}{2}
\sigma }.  \label{trans3}
\end{equation}
The relevant point is that after the reparametrization of the scalar $\phi
=\chi ^{\frac{4-n}{2-n}}$ the form of the action (\ref{conf2n}) changes, and
the transformation rule for the new field $\chi $ is exactly the same as for
the scalar field $\varphi $ of the theory (\ref{conf1}). Indeed, the above
reparametrization becomes degenerate at $n=4$. Thus in any dimension
$n\neq 4$
there are two different conformal models (\ref{conf2n}) and (\ref{conf1n})
which include the second $\Delta _{2}^{(n)}$ and the fourth $\Delta
_{4}^{(n)}$ order operators in the dilaton sector. For any $n\neq 4$, with
accuracy to the parametrization of the scalar field, the field content and
symmetries of the theories are exactly the same. However we fail to see the
transformation which links two conformal models. On the other hand, another
kind of relation between them arises if we consider both models as a
one-loop renormalizable versions of the general high derivative theory (\ref
{gener}). We shall discuss quantum aspects of the theory in the next
section.

\vskip 6mm 
\noindent {\large {\bf 3. Renormalization of second conformal model in
arbitrary parametrization.}}
\vskip 2mm

The purpose of this section is to investigate the one-loop renormalization
of the theory (\ref{conf2}). As we already know, for the classical theory
the form of the function $\zeta (\phi )$ in the action can be changed by
reparametrizing the scalar field. If we intend to study the one-loop
effective action in the framework of the background field method, then the
reparametrization of the background field doesn't create anything new with
respect to the classical theory. On the other hand, the reparametrization of
the quantum field can essentially affect the one-loop divergencies, so that
only the on-shell quantities remain invariant\footnote{%
The higher loop contributions can depend on the parametrization in a
more complicated way, because in this case one has to take into account the
quantum corrections to the equations of motion \cite{param,book}.}. In our
case, due to its simplicity, the theory (\ref{conf2}) gives an opportunity
to trace the off-shell dependence on the parametrization of the quantum
field explicitly in a general form (we remind that only the scalar is a
quantum field). To do this one has to write the action (\ref{conf2}) as the
particular form of (\ref{gener}) and then use the result for the one-loop
divergences of the last theory derived in \cite{eli}. The problem is to find
an explicit form of the renormalization for the function $\zeta (\phi )$ and
for the parameters $v,x,y,z,u$, which removes the counterterms in an
arbitrary parametrization.

The conformal invariant action (\ref{conf2}) corresponds to the following
constraints for the functions
$\,a_{i}(\phi ),\,b_{j}(\phi ),\,c_{k}(\phi )\,$
of the model (\ref{gener}), where we put $\ga = 1$ for simplicity.
\[
b_{1}=\frac{ y(\zeta^\prime )^2}{\zeta^2}
,\,\,\,\,\,\,\,\,\,\,\,
\,b_{2} = \frac {y\zeta^\prime \,
(2\,\zeta \zeta ^{\prime\prime} - (\zeta^\prime)^2) }{\zeta ^3}
,\,\,\,\,\,\,\,\,\,\,\,\,
b_{3}=y\,
\frac{ 4\,\zeta ^{2}\left( \zeta ^{\prime \prime }\right)^2 +
\left( \zeta^{\prime }\right)^4 - 4\,\zeta \zeta ^{\prime \prime}
(\zeta ^{\prime })^2}{4\zeta^4}
\]
\[
b_{4}=\frac{3}{2}\,{\frac{z(\zeta ^{\prime })^{2}}{\zeta }}%
,\,\,\,\,\,\,\,\,\,\,\,\,b_{5}=t\zeta ^{2},\,\,\,\,\,\,\,\,\,\,\,\,c_{1}=
y{\frac{(\zeta ^{\prime })^{2}-
2\,\zeta \zeta ^{\prime \prime }}{3\,\zeta ^{2}}
},\,\,\,\,\,\,\,\,\,\,\,\,c_{2}=0
,\,\,\,\,\,\,\,\,\,\,\,\,
c_{3}=-\frac23\,{\frac{y\,\zeta ^{\prime }}{\zeta }}
\]
\begin{equation}
a_{1}=x+v,\,\,\,\,\,\,\,\,\,\,a_{2}=2\,x-4\,v,\,\,\,\,\,\,\,\,\,\,a_{3}=
-\frac{1}{3}\,x+v+\frac{y}{9},\,\,\,\,\,\,\,\,\,\,a_{4}=z\zeta .  \label{constr2}
\end{equation}

The one-loop counterterms for the general model (\ref{gener}) have the form: 
\[
\Delta S={\frac{\mu }{\varepsilon }}^{n-4}\int d^{n}x\sqrt{-g}\,\left\{
A_{1}R_{\alpha \beta \mu \nu }^{2}+A_{2}R_{\alpha \beta
}^{2}+A_{3}R^{2}+A_{4}R + C_{1}R(\nabla \varphi )^{2}+C_{2}R^{\alpha \beta
}\,\partial _{\al}\varphi \,\partial _{\be}\varphi +\right. 
\]
\begin{equation}
\left. +C_{3}R(\Box \varphi )+B_{1}(\Box \varphi )^{2}+B_{2}(\nabla
{\varphi })^{2}(\Box \varphi )+B_{3}(\nabla \varphi )^{4}+B_{4}(\nabla
\varphi )^{2}+B_{5}\right\} ,  \label{countr}
\end{equation}
where $\varepsilon =(4\pi )^{2}(n-4)$ and cumbersome explicit expressions
for the coefficients $A_{i},\,B_{j},\,C_{k}\,$ can be found in \cite{eli}
\footnote{
The expression for $B_{4}$ in \cite{eli} contains a misprint. The correct
form is 
\[
B_{4}=-\frac{b_{4}^{\prime \prime }}{2b_{1}}+\frac{1}{2b_{1}^{2}}
(4b_{1}^{\prime \prime }b_{4}-4b_{2}^{\prime
}b_{4}+6b_{3}b_{4}+5b_{1}^{\prime }b_{4}^{\prime }-3b_{4}^{\prime }b_{2})
+\frac{1}{b_{1}^{3}}(3b_{1}^{\prime }b_{2}b_{4}-
4(b_{1}^{\prime })^{2}b_{4})
\]
}. After substituting the constraints (\ref{constr2}) into (\ref{countr})
we obtain the following expressions for the counterterms. 
\[
B_{1}=\frac{2(\zeta ^{\prime \prime })^{2}}{(\zeta ^{\prime })^{2}}-
\frac{\zeta ^{\prime \prime }}{(\zeta ^{\prime })^{2}\zeta }-
\frac{{\zeta ^{\prime
\prime \prime }}}{\zeta ^{\prime }}+
\frac{5}{4}\frac{(\zeta ^{\prime })^{2}}{\zeta ^{2}},
\]
\[
B_{2}=\frac{5\zeta ^{\prime \prime }{\zeta ^{\prime \prime \prime }}}{
(\zeta ^{\prime })^{2}} 
- \frac{5{\zeta ^{\prime \prime \prime }}}{2\zeta } +
\frac{3(\zeta ^{\prime \prime })^{2}}{\zeta ^{\prime }\zeta }
+\frac{2\zeta^{\prime }}{\zeta ^{2}}
-\frac{\zeta ^{\prime \prime \prime \prime }}{\zeta^{\prime }}
-\frac{5\,(\zeta ^{\prime })^{3}}{4\,\zeta ^{3}}
-{\frac{4\,(\zeta^{\prime \prime })^{3}}{(\zeta ^{\prime })^{3}\,}}\,,
\]
\begin{eqnarray*}
B_{3} &=&\frac{1}{16}\left\{ \frac{-96\,\zeta ^{\prime \prime \prime }\zeta
^{\prime \prime }}{\zeta ^{\prime }\zeta }-
\frac{16\,\zeta ^{\prime \prime
}\zeta ^{\prime \prime \prime \prime }}{(\zeta ^{\prime })^{2}}+
\frac{24\,\zeta ^{\prime }\zeta ^{\prime \prime \prime }}{\zeta ^{2}}+
\frac{8\,\zeta ^{\prime \prime \prime \prime }}{\zeta }-
\frac{36(\zeta ^{\prime\prime })^{2}}{\zeta ^{2}}\right.  \\
&&\left. +\frac{5(\zeta ^{\prime })^{4}}{\zeta ^{4}}-\frac{12\,\zeta
^{\prime \prime }(\zeta ^{\prime })^{2}}{\zeta ^{3}}+\frac{128(\zeta
^{\prime \prime })^{3}}{(\zeta ^{\prime })^{2}\zeta }+\frac{96\,\zeta
^{\prime \prime \prime }(\zeta ^{\prime \prime })^{2}}{(\zeta ^{\prime })^{3}
}-{\frac{96\,(\zeta ^{\prime \prime })^{4}}{(\zeta ^{\prime })^{4}}}\right\}
\,,
\end{eqnarray*}
\[
B_{4}=\frac{3}{8}{z\left( \frac{8\zeta ^{2}(\zeta ^{\prime \prime
})^{2}-6\zeta \zeta ^{\prime \prime }(\zeta ^{\prime })^{2}-4\zeta ^{\prime
}\zeta ^{\prime \prime \prime }\zeta ^{2}-3(\zeta ^{\prime })^{4}}{y(\zeta
^{\prime })^{2}\zeta }\right) ,}
\]
\[
B_{5}={\frac{9}{8}}\,{\zeta ^{2}\left( \frac{z^{2}}{y^{2}}
-\frac{8\,t(\zeta ^{\prime })^{2} + 8\,t\zeta \zeta ^{\prime \prime }}{
y(\zeta ^{\prime })^{2}}\right) ,}
\]
\[
C_{1}=\frac{5\,(\zeta ^{\prime })^{2}}{12\,\zeta ^{2}}-\frac{5(\zeta
^{\prime \prime })^{2}}{3\zeta ^{2}(\zeta ^{\prime })^{2}}-\frac{\zeta
^{\prime \prime }}{2\zeta }+\frac{\zeta ^{\prime \prime \prime }}{\zeta
^{2}\zeta ^{\prime }}+\frac{\zeta \zeta ^{\prime \prime \prime \prime }}{
3(\zeta ^{\prime })^{2}}-\frac{2\zeta \zeta ^{\prime \prime }\zeta ^{\prime
\prime \prime }}{(\zeta ^{\prime })^{3}}+\frac{2\zeta (\zeta ^{\prime \prime
})^{3}}{(\zeta ^{\prime })^{4}},
\]
\[
C_{2}=0
,\,\,\,\,\,\,\,\,\,\,\,\,\,\,\,
C_{3}=\frac{1}{6}\left\{ \frac{%
2\,\zeta^{\prime \prime }}{\zeta ^{\prime }}-\frac{5\,\zeta
^{\prime }}{\zeta }+\frac{2\,\zeta ^{\prime \prime \prime }\zeta }{(\zeta
^{\prime })^{2}}-\frac{4\,\zeta (\zeta ^{\prime \prime })^{2}}{(\zeta
^{\prime })^{3}}\,\right\} ,
\]
\begin{equation}
A_{1}={\frac{1}{90}},\,\,\,\,\,\,\,\,\,\,A_{2}=-{\frac{1}{90}}%
,\,\,\,\,\,\,\,\,\,\,A_{3}={\frac{5}{36}},\,\,\,\,\,\,\,\,\,\,A_{4}=-{\frac{
z\zeta \left( 2\zeta \zeta ^{\prime \prime }+3(\zeta ^{\prime })^{2}\right) 
}{4\,y\,(\zeta ^{\prime })^{2}}.}  \label{coeff}
\end{equation}

According to general theorems \cite{param} (see also \cite{book}), the above
formulas have to meet two requirements:

i) The coefficients $A_{i},\,B_{j},\,C_{k}\,$ satisfy the conformal
constraints (\ref{constr2}).

ii) The divergences of the renormalized action $S+\Delta S$ can be removed
by the renormalization of the function $\zeta$ and of the constant
parameters $x,y,z,u$.

Indeed, the transformation of the function $\,\,\zeta _{0}=\zeta
+(1/\varepsilon )f(\zeta ,\phi )\,\,$ can be always solved with respect
to $\phi $ with accuracy to ${\cal O}(1/\varepsilon )$, and therefore the
renormalization of $\zeta $ is nothing but another form of the 
renormalization of the field
$ \phi_0 = \phi + \frac{1}{\varepsilon}\,f(\ze)/\ze^{prime}$. 
Of course, for general $\zeta $ such a
renormalization can be nonlinear and quite complicated. The same
transformation
can be viewed from a different point of view. As we already
know
from the previous section, any change of parametrization for the scalar
field doesn\'{}t
violate the conformal invariance. It only modifies the form
of the conformal transformation (\ref{trans}). Therefore,  i) follows from
ii), and we have to check only the former.

The search of the renormalization transformation ii) is not easy because of
the complicated form of the expressions (\ref{coeff}). That is why we shall
construct such a transformation in two steps. First we take a simple
particular form of $\zeta =\phi ^{\al}$, and find the form of the
renormalization
constants for the parameters $v,x,y,z,u$. After that we find
an explicit form of the renormalization for the function $\zeta (\phi )$.

For the special case $\zeta =\phi ^{\al}$ the expressions (\ref{coeff}) have
the following form. First, 
\begin{equation}
(B_{1},B_{2},B_{3},C_{1},C_{2},C_{3})=\frac{5}{4y}%
\,(b_{1},b_{2},b_{3},c_{1},c_{2},c_{3}),  \label{1}
\end{equation}
while $b_{1},b_{2},b_{3},c_{1},c_{2},c_{3}$ can be all restored from the
constraints (\ref{constr2}). Say, $b_{1}=y\alpha ^{2}\phi ^{-2}$ etc. This
result indicates that for this special parametrization the conformal
invariance in the high derivative sector of the model is preserved, that the
renormalization of the function $\zeta (\phi )$ is not necessary, and that
in the high derivative sector the renormalization constant for $y$ is
parametrization independent. The last property have to hold for the general
parametrization too. Since the relation between the above parameters is
fixed, it is sufficient to consider the renormalization of one parameter,
for example $b_{1}$. 
All others will renomalize identically. For the second- and
zero-derivative sectors of the model we find 
\[
a_{4}=z\phi ^{\al},\,\,\,\,\,\,\,\,\,\,\,b_{4}=\frac{3}{2}\,\alpha
^{2}\,\phi ^{-2}\,a_{4},\,\,\,\,\,\,\,\,\,\,\,b_{5}=u\phi ^{2\alpha },
\]
\begin{equation}
A_{4}=-\frac{1}{4}\,
\frac{z}{y}\,\left( 5-\frac{2}{\alpha }\right) \phi ^{\al}
,\,\,\,\,\,\,\,\,\,\,\,B_{4}=\frac{3}{2}\,
\alpha ^{2}\,\phi^{-2}\,A_{4}
,\,\,\,\,\,\,\,\,\,\,\,B_{5}=\left( \frac{1}{\alpha }\,
\frac{u}{y}-2\,\frac{u}{y}+\frac{9}{8}\,\frac{z^{2}}{y^{2}}\right) \,
\phi ^{2\alpha }.
\label{second}
\end{equation}
The last formula indicates that 
the lower order counterterms are conformal invariant, and the relation
between $A_{4}$ and $B_{4}$ is fixed.
All the counterterms can be removed by the renormalization of the
parameters 
$p=(v,x,y,z,u)$
\begin{equation}
p_{0}=\mu ^{n-4}\,\left( p+\frac{F_{p}}{\varepsilon }\right) .
\label{renorm}
\end{equation}
One can perform an additional renormalization of the scalar
$\phi _{0}=\phi -
\frac{1}{\varepsilon }\,\frac{\beta }{y}$ with an arbitrary $\beta $, and
similar (global) renormalization of the metric. As we know from the previous
section, the effect of such renormalization is the same as the one for the
scalar field, but we include this transformation $g_{\mu \nu }^{0}=g_{\mu
\nu }\,\left( 1-\frac{1}{\varepsilon }\,\frac{\gamma }{y}\right) $ ,
with an
arbitrary parameter $\gamma $, for the sake of completeness.

Taking all this into account, the renormalization parameters become 
\begin{equation}
F_{v}=-\frac{1}{180},,\,\,\,\,\,\,\,\,\,\,\,\,\,\,\,\,\,\,\,\,\,\,
F_{x}=\frac{1}{60}
,\,\,\,\,\,\,\,\,\,\,\,\,\,\,\,\,\,\,\,\,\,\,F_{y}=-\frac{5}{4},
\label{ren2}
\end{equation}
\begin{equation}
F_{z}=\left( \frac{1}{2\alpha }-\frac{5}{4}+\alpha \beta +\gamma \right) 
\frac{z}{y}
,\,\,\,\,\,\,\,\,\,\,\,\,\,
F_{u}=\left( \frac{1}{\alpha }-2+2\alpha \beta +2\gamma \right) 
\frac{u}{y}+\frac{9}{8}\,\frac{z^{2}}{y^{2}%
}.  \label{ren3}
\end{equation}
One can see that the renormalization constants for $v,x,y$ don't, but the
renormalization constants for $z$ and $u$ do depend on the function
$\zeta $,
that is on the parametrization of the quantum scalar field. The two
quantities $F_{z},F_{u}$ depend also on an arbitrary parameters $\beta
,\gamma $. One can, indeed, choose $\beta $ and $\gamma $ in such a way that
both $F_{z},F_{u}$ become independent on $\alpha $. For example, this can be
achieved for $\beta =-\frac{1}{2\alpha ^{2}},\,\,\gamma =0$. However this
doesn't mean that we are able to provide the unambiguous renormalization of
the individual parameters $z,u$, because such a choice of $\beta ,\gamma $
is not better than any other one. The crucial observation is that the
combination $2uF_{z}-zF_{u}$ is parametrization independent. This
combination
is nothing but the renormalization parameter for $\frac{z^{2}}{t}
$ which is an essential coupling of the theory.

Now we are in the position to study the one-loop renormalization of the
theory (\ref{conf2}) in an arbitrary parametrization. Taking into account
the renormalization of $y$ in (\ref{renorm}), (\ref{ren2}) we were able to
establish the transformation for $\zeta (\phi )$ which, together with the
``high-derivative'' part of (\ref{renorm}) and (\ref{ren2}), provides the
finiteness in the corresponding sector of the theory. 
\begin{equation}
\zeta ^{(0)}\,=\,\mu ^{n-4}\,\left\{\, \zeta\, -
\,\frac{1}{2\,\varepsilon }\,\,\zeta
^{\prime \prime }\,
\left( \frac{\zeta ^{\prime }}{\zeta }\right)^{2}\,\right\} .
\label{great}
\end{equation}
Direct substitution shows that the above transformation, together with the
appropriate renormalization of $v,x,y,z,u$ removes all the counterterms. The
renormalization of the parameters $v,x,y,z,u$ has the form (\ref{renorm}),
(\ref{ren2}),(\ref{ren3}) with any particular values which satisfy
$\,\,\ga + \alpha \beta =\frac{\alpha -1}{2\alpha },\,$.
Indeed, this doesn't mean that the
renormalization of the parameters $z,u$ is independent on the
parametrization of the quantum field. To ensure in the opposite it is
enough
to remind the results for the particular parametrization
$\zeta =\phi ^{\al}$,
which we have discussed above. Only the renormalization of the
dimensionless ratio $u/z^{2}$ is invariant and unique.

Thus, the theory under consideration is one-loop renormalizable. Contrary to
the first conformal model (\ref{conf1}) explored in \cite{eli}, our model
(\ref{conf2})
doesn't allow finite solutions. This follows directly from the
existence of the parameters $y$ and $u/z^{2},$ which have universal and
non-zero renormalization. Therefore, in this model one always meets the
conformal anomaly, which modifies the finite part of the effective action,
and breaks the conformal symmetry.

\vskip 6mm 
\noindent {\large {\bf 3. Renormalization group equations}} \vskip 2mm

In the previous section we have seen that the theory (\ref{conf2}) is
one-loop renormalizable, so we can use the renormalization group method to
investigate its scaling behaviour. Here we regard (\ref{conf2}) as a theory
of quantum scalar field in curved spacetime, and therefore the
renormalization group should also be formulated in curved spacetime (one can
look at \cite{book} for the introduction and references on the
renormalization group in an external gravitational field). The
renormalization group equation for the effective action has the following
solution,
\begin{equation}
\Gamma \left[ e^{-2t}\,g_{\mu \nu },\phi ,p,\mu \right] =\Gamma \left[
g_{\mu \nu },\phi (t),p(t),\mu \right] ,  \label{URG}
\end{equation}
which enables one to explore the short-distance (and therefore high-energy)
limit, which we shall call UV. In the massless conformal theories the same
equation can serve for the investigation of the long-distance (IR) limit.
The scaling behaviour of the effective charges $p(t)=\left(
x(t),y(t),z(t),u(t),v(t)\right)$ 
is related to the conventional $\beta $-functions. 
\begin{equation}
\beta _{p}(n)=\mu \frac{dp}{d\mu },  \label{beta}
\end{equation}
where the limit $\beta _{p}=\lim_{n\rightarrow 4}\,\beta _{p}(n)$ should be
taken after the operation (\ref{beta}) over the renormalized parameter $p$.
One can construct also the renormalization group equations for 
$\phi (t)$ and $g_{\mu \nu }(t)$ but they will be ambiguous, and we
do not write them here.
For the sake of simplicity we present the expressions for the $\beta $
-functions for the spacial parametrization $\zeta =\phi ^{\al}$. As it was
demonstrated in the previous section, the renormalization of the scalar
field and metric depend on the arbitrary parameters $\beta $ and
$\gamma $
(\ref{ren3}). Therefore the corresponding renormalization group
equations do
not have independent physical sense and we write them only in order to
illustrate the parametrization dependence \footnote{
This high degree (two-parametric) of ambiguity is in fact caused by the
conformal (global) invariance of the theory, it can be weakened if one
introduces the mass of the scalar field or Einstein term.}. The
renormalization group equations for $x(t),y(t),z(t),u(t),v(t)$ have the form 
\[
(4\pi )^{2}\,\frac{dv}{dt}=-\varepsilon \,v+
\frac{1}{180},\,\,\,\,\,\,\,\,\,%
\,\,\,\,\,v(0)=v_{0},
\]
\[
(4\pi )^{2}\,
\frac{dx}{dt}=-\varepsilon \,x-\frac{1}{60},\,\,\,\,\,\,\,\,\,
\,\,\,\,\,x(0)=x_{0},
\]
\[
(4\pi )^{2}\,\frac{dy^{-1}}{dt}=\varepsilon \,y^{-1}+\frac{5}{4}
\,y^{-2},\,\,\,\,\,\,\,\,\,\,\,\,\,\,y(0)=y_{0},
\]
\[
(4\pi )^{2}\,\frac{dz}{dt}=-\varepsilon \,z+\left( \frac{5}{4}-
\frac{1}{2\alpha }-
\alpha \beta -\gamma \right) \,\frac{z}{y},\,\,\,\,\,\,\,\,\,\,\,
\,\,\,z(0)=z_{0},
\]
\begin{equation}
(4\pi )^{2}\,\frac{du}{dt}=-\varepsilon \,u+\left( 2-\frac{1}{\alpha }
-2\alpha \beta -2\gamma \right) \,\frac{u}{y}-
\frac{9}{8}\,\frac{z^{2}}{y^{2}%
},\,\,\,\,\,\,\,\,\,\,\,\,\,\,u(0)=u_{0},  \label{RGE}
\end{equation}
where $\varepsilon =(4\pi )^{2}(n-4)$. We have written the renormalization
group equation for $y^{-1}$ because just this quantity is the parameter of
the loop expansion in the path integral, so the behaviour of $y^{-1}$ in UV
and IR limits defines the asymptotical properties of the theory. The
equations for the individual $z(t),u(t)$ are ambiguous, so that one can
define uniquely only the behaviour of the essential coupling constant
$\lambda =\frac{u}{z^{2}}$, for which we have 
\begin{equation}
(4\pi )^{2}\,\frac{d\lambda }{dt}=\varepsilon \,\lambda -\frac{1}{2y}
\,\lambda -\frac{9}{8\,y^{2}},\,\,\,\,\,\,\,\,\,\,\,\,\,\,\,\,
\la(0)=\frac{%
u_{0}}{z_{0}^{2}}  \label{RGE2}
\end{equation}

The solution of the above equations is straightforward. We present the
solutions for $n=4$. 
\[
x(t)=x_{0}-\frac{1}{60\,(4\pi )^{2}}\,t
,\,\,\,\,\,\,\,\,\,\,\,\,\,\,\,%
\,v(t)=v_{0}+\frac{1}{180\,(4\pi )^{2}}
\,t,\,\,\,\,\,\,\,\,\,\,\,\,\,\,\,\,
y^{-1}(t)=y_{0}^{-1}
\left( 1-\frac{5}{4(4\pi )^{2}}\,y_{0}^{-1}t\right)^{-1},
\]
\begin{equation}
\lambda (t)=\left( \lambda _{0}+\frac{9}{14\,y_{0}}\right) \,
\left( \frac{y(t)}{y_{0}}\right) ^{\frac{2}{5}}-\frac{9}{14\,y(t)}.
\label{solu}
\end{equation}
The solution for $y(t)$ indicates that the theory (\ref{conf2}) is
asymptotically free in IR. The formulas for $x(t),v(t)$ describe linear
variations of these effective charges with scale. The last solution gives
the evolution of the cosmological constant measured in units of the
Newtonian constant (or {\it v.v.}). For the spacial initial conditions of
the renormalization group flow $\lambda _{0}=-\frac{9}{14\,y_{0}}$ we meet
the decreasing cosmological constant at low energies. We remark that the
behaviour of the cosmological $u$ and gravitational $z$ constants is
ambiguous, as it always happens. In our case this ambiguity is
caused by the
freedom to change quantum variables -- reparametrizations in the scalar
sector -- or by the freedom to reparametrize 
the conformal factor of the metric.

\vskip 6mm 
\noindent {\large {\bf 4. Conclusions}} \vskip 2mm

We have investigated some classical and quantum aspects of the second
(alternative) conformal limit of the recently proposed general higher
derivative dilaton quantum theory in curved spacetime. In particular, the
form of conformal transformation for two conformal metric-dilaton models
with fourth derivatives has been compared, and it was shown that in the
spacetime dimension $n\neq 4$ the transformations lows are identical with
accuracy to the reparametrization in the scalar sector. We also generalize
the conformal duality of \cite{bek,conf,duco} for the high derivative
metric-dilaton theories.

The quantum theory of dilaton on an arbitrary metric background was
investigated. Using the result of our earlier calculations \cite{eli} it
was
demonstrated that the second conformal invariant model is multiplicatively
renormalizable in one loop, while for the first conformal limit the same is
known from \cite{eli}. The renormalization of the dimensionless parameters
was shown to be universal for an arbitrary parametrization of the quantum
field. The simplicity of the models enables one to trace such a
parametrization dependence explicitly, thus we find a compensating
transformation for the background field. The renormalization group
equations
for the essential coupling constants indicate an asymptotic freedom in the
IR limit. In this respect, the theory is similar to the well-known model of
Antoniadis and Mottola \cite{anmo}, based on the anomaly-induced effective
action.

In order to outline the possible directions for future studies, we can
mention the possibility to investigate the renormalization, renormalization
group, and specially the renormalization group improved effective potential
for the theory with conformal duality (\ref{dual2}). This is a potentially
important problem, because the conformal duality can link the regimes of
strong and weak gravitational field, and, therefore, we may hope to get an
information about the effective potential in the vicinity of the
singularities of a black hole and Big-Bang solutions. Indeed the relevance
of such investigation showld be considerably higher if we would be able to
make the calculations in the full theory with a quantum metric and scalar.
The method of calculation in such a theory is known \cite{a}, but the actual
computations seems, at the moment, unfeasable to us, due to its algebraic
complexity.
Alternatively, we can use some approximate methods like the $1/N$
expansion, adding to the original theory a new set of $N$ conformal matter
fields. Another interesting possibility is to solve the equation (\ref
{effact}) with the anomalous trace of the energy-momentum tensor which
corresponds to the higher derivative conformal scalar models (\ref{conf1})
and (\ref{conf2}).
The {\it r.h.s.} of (\ref{effact}) is proportional to the
one-loop divergences for this theories \cite{duff-77}, which are all known
from \cite{eli} and the present paper. The corresponding solution of (\ref
{effact}) gives the finite quantum correction to the action of the theory.
Therefore, it can be very interesting, in view of possible cosmological
applications, because for conformally flat cosmological metrics the
integration constant for (\ref{effact}) is irrelevant. One can mention, for
instance, that a similar solution with torsion leads to inflationary model 
\cite{buodsh}. We hope to investigate the above problems in the close
future.

\vskip 6mm 
\noindent {\large {\bf Acknowledgements}} \vskip 2mm

One of the authors (I.Sh.) is grateful to the Departamento de Fisica, UFJF
for warm hospitality. The work of I.Sh. was supported in part by Russian
Foundation for Basic Research under the project No.96-02-16017.  J.A.B.
thanks FAPEMIG (Minas Gerais State Sponsoring Agency) for finantial support. 



\vskip 20mm

\begin{thebibliography}{99}
\bibitem{20let}  M.J. Duff, Class.Quant.Grav. {\bf 11}, 1387 (1994).

\bibitem{odsh}  S.D. Odintsov and I.L. Shapiro, Class. Quant. Grav. {\bf 8}
L57 (1991).

\bibitem{anmo}  I. Antoniadis and E. Mottola, Phys. Rev. {\bf 45D}, 2013
(1992).

\bibitem{desc}  S. Deser and A. Schwimmer, Phys.Lett. {\bf 309B} 279 (1993).

\bibitem{cosh} I.L. Shapiro and G. Cognola, Phys.Rev. {\bf 51D} 2775 (1995).

\bibitem{deser}  S. Deser, Helv.Phys.Acta {\bf 69} (1996) 570,
hep-th/9609138.

\bibitem{rei}  R.Y. Reigert, Phys.Lett. {\bf 134B}, 56 (1984).

\bibitem{frts}  E.S. Fradkin and A.A. Tseytlin, Phys.Lett. {\bf 134B}, 187
(1984).

\bibitem{buodsh}  I.L. Buchbinder, S.D. Odintsov and I.L. Shapiro,
Phys.Lett. {\bf 162B}, 92 (1985).

\bibitem{bugufo}  I.L. Buchbinder, V.P. Gusynin and P.I. Fomin, Yad. Fiz.
(Sov. J. Nucl. Phys.) {\bf 44}, 828 (1986).

\bibitem{dow}  J.S. Dowker, Phys.Rev. {\bf D33}, 3150 (1986).

\bibitem{osbpet}  H. Osborn and A. Petkou, Ann.Phys. {\bf 231} (1994) 311.

\bibitem{erdosb}  J. Erdmenger and H. Osborn, Nucl.Phys. {\bf 483} (1996)
431.

\bibitem{vilk}  G.A. Vilkovisky, Class.Quant.Grav. {\bf 9} 895 (1992); A.O.
Barvinsky, Yu. V. Gusev, G.A. Vilkovisky and Zhytnikov, J.Math.Phys.
{\bf 35} 3525;3543 (1994);
Nucl.Phys. {\bf B 439} 561 (1995); A.O. Barvinsky, A.G.
Mirzabekian and V.V. Zhytnikov, gr-qc/9510037.

\bibitem{eli}  E. Elizalde, A.G. Jacksenaev, S.D. Odintsov, I.L.Shapiro,
Class.Quant.Grav. {\bf 12} 1385 (1995); Phys.Lett. {\bf 328B} 297 (1994);
E.Elizalde, S.D. Odintsov and I.L. Shapiro, Class.Quant.Grav. {\bf 11} 1607
(1994).

\bibitem{anilto}  I. Antoniadis, J. Iliopoulos and T.N. Tomaras, Nucl.Phys. 
{\bf B261}157 (1985);

\bibitem{conf}  I.L. Shapiro and H. Takata, Phys.Lett. {\bf 361 B}, 31
(1996).

\bibitem{bek}  J.D. Bekenstein, Ann.Phys. {\bf 82},535,(1974).

\bibitem{duco}  I.L. Shapiro, Class.Quant.Grav. {\bf 14}, 391 (1997).

\bibitem{a}  I.L. Shapiro and A.G. Jacksenaev, Phys.Lett. {\bf 324B}, 284
(1994).

\bibitem{param}  B.L. Voronov, P.M. Lavrov and I.V. Tyutin, Sov.J.Nucl.Phys. 
{\bf 36} (1982) 498;
I.V. Tyutin, Sov.J.Nucl.Phys. {\bf 35} (1982) 125.

\bibitem{stelle}  K.S. Stelle, Phys.Rev. {\bf 16D}, 953 (1977).

\bibitem{buch}  I.L. Buchbinder, Theor.Math.Phys. {\bf 61}, 393 (1984).

\bibitem{book}  I.L. Buchbinder, S.D. Odintsov and I.L. Shapiro, {\sl 
Effective Action in Quantum Gravity} (IOP, Bristol, 1992).

\bibitem{hajo}  H.-J. Schmidt, {\sl A new duality transformation for
fourth-order gravity}, hep-th/9703002, Gen.Rel.Grav., to be published.

\bibitem{barrow}  J.D. Barrow, S. Cotsakis, Phys.Lett. {\bf 214B}, (1988)
515.

\bibitem{pan}  S. Paneitz, A Quartic Conformally Covariant Differential
Operator for Arbitrary Pseudo-Riemannian Manifolds, MIT preprint, 1983.

\bibitem{branson}  T.P. Branson,
Comm.Part.Diff.Equations {\bf 7} (1982)
393; Math.Scand. {\bf 57} (1985) 293.

\bibitem{branson2} T.P. Branson,
 Commun.Math.Phys. {\bf 178} (1996) 301.

\bibitem{erd}  J. Erdmenger, hep-th/9704108.

\bibitem{duff-77}  M.J. Duff, Nucl.Phys. {\bf B125}, (1977) 334.
\end{thebibliography}
\end{document}